\documentclass[aps,apl,groupedaddress,twocolumn]{revtex4-1}
\usepackage{graphicx}
\usepackage{amsmath}
\usepackage{epstopdf}

\begin{document}


\title{Self-Assembled Graphene Plasmon Resonators}


\author{}
\affiliation{}


\date{\today}

\begin{abstract}
The ability to fabricate dense small features over a large area is important for graphene plasmonics. We present the first self-assembled graphene plasmonic resonators operating in the mid-IR. The resonators are 35 nm in diameter with 20 nm spacing and cover a centimeter sized area. The resonators exhibit a very broad resonance. We fit our data using a drude model and combine it with SEM data to investigate the contribution to broadening from our process. We find that the self-assembly does not contribute significantly to the broadening observed.  
\end{abstract}

\pacs{}

\maketitle

The Terahertz regime is of interest for sensing applications because many chemical compounds possess a unique fingerprint in the IR region\cite{THzSensing}. One promising route to components operating in the Terahertz is by utilizing graphene plasmons because of the high field effect mobility\cite{Petrone2012} and tunable conductivity of graphene\cite{FWang2011}. However studying graphene plasmons in the mid-IR region has been challenging due to the requirement of several tens of nanometer sized features over centimeter sized areas. In this regard several groups have employed expensive electron beam lithography or THz objectives which are lossy to get around this challenge\cite{Hugen2013,Koppens,Brar,Xiangfeng}. In terms of applications these techniques are not scalable, therefore there is a need to investigate alternative technologies. In this work we introduce a novel process to produce graphene plasmon resonators over large area using block copolymer self-assembly\cite{Chuck2012}. We measured our devices using FTIR spectroscopy and confirm our data by fitting it to a drude model. In the second part of the paper we investigate the effect our process has on the quality factor of our graphene plasmon resonators.
\begin{figure}[h]
\includegraphics[scale=0.38]{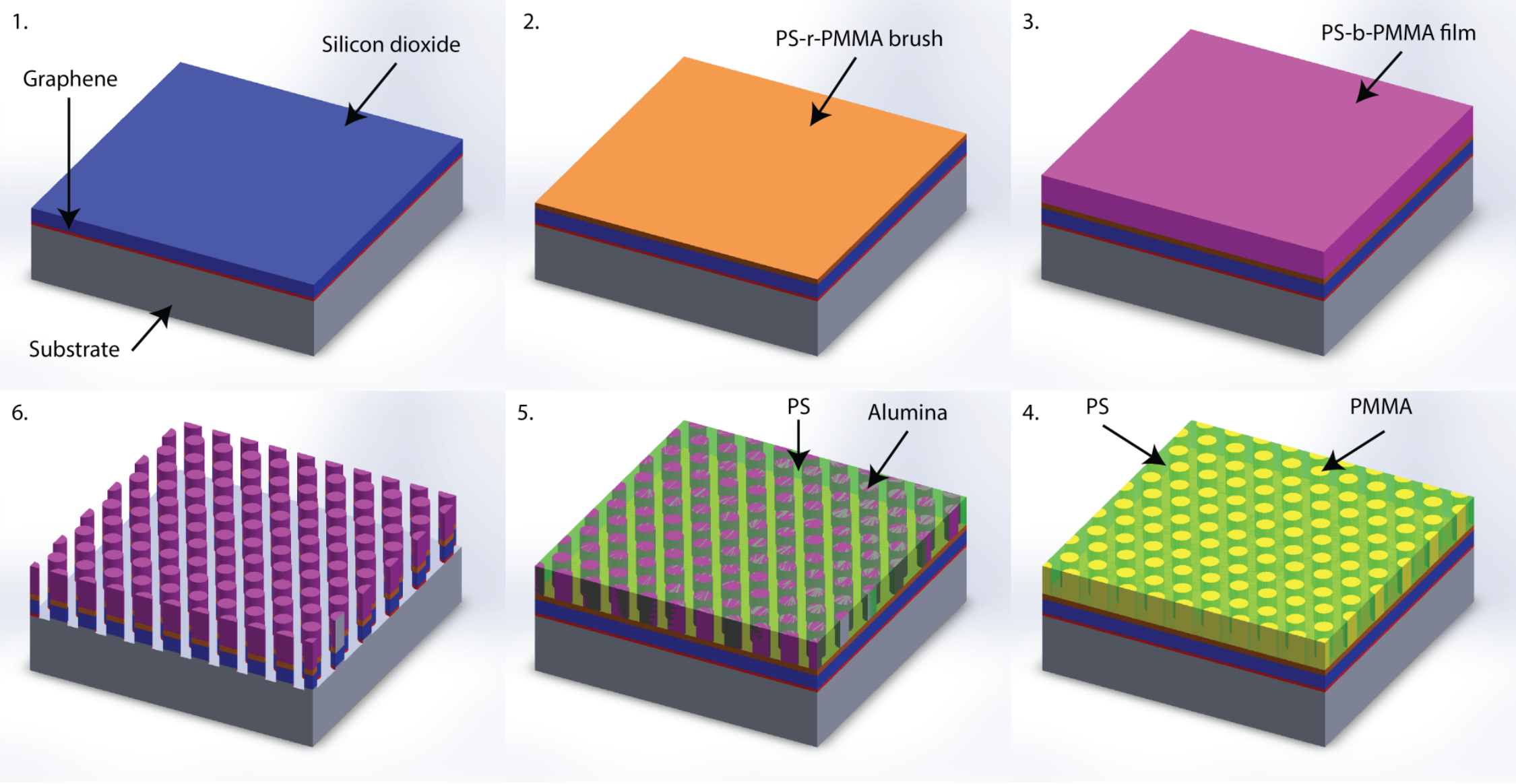}
\caption{Process flow for self-assembled graphene resonators}
\end{figure}

Fig 1 shows the process flow. Graphene is grown on CVD copper and transferred to 90 nm SiO2 on a double side polished high resistivity substrate. The transferred graphene is then dipped in 50\% Nitric Acid for 60 sec and dried without further rinsing. Next 8 nm of SiO$_{2}$ is evaporated on to the graphene at 0.3A/s in 10$^{-6}$ Torr vacuum. A 2\% random copolymer in toluene from polymer source is spun coat at 2000 rpm on to the SiO$_{2}$ and annealed overnight for 12 hours in vacuum at 180C to form a brush. At each step of the process we clean the backside of our sample using acetone to ensure we can obtain a high enough signal during our measurements. After that the excess brush is rinsed off by spinning the substrate at 2000 rpm and spraying toluene on to the brush. The substrate is then blow dried using nitrogen and a 2\% PS-b-PMMA is spun coat onto the brush at 2000 rpm.  The PS-b-PMMA film stack is then annealed for 12 hours overnight in vacuum at 180C. During annealing the PS and PMMA domains phase separate to form perpendicularly aligned cylinders. We infuse the cylinders with alumina in an ALD chamber since the alumina preferentially nucleates on the PMMA domains. Finally using the alumina pillars as an etch mask we etch through the PS scaffold and oxide/graphene stack using CF$_{4}$/O$_{2}$ RIE to produce a substrate with 35 nm graphene dots of 20 nm spacing. We performed SEM microscopy to demonstrate the uniform distribution of the dots over large areas as shown in Figure 2.
\begin{figure}[h]
\includegraphics[angle=0,scale=0.28]{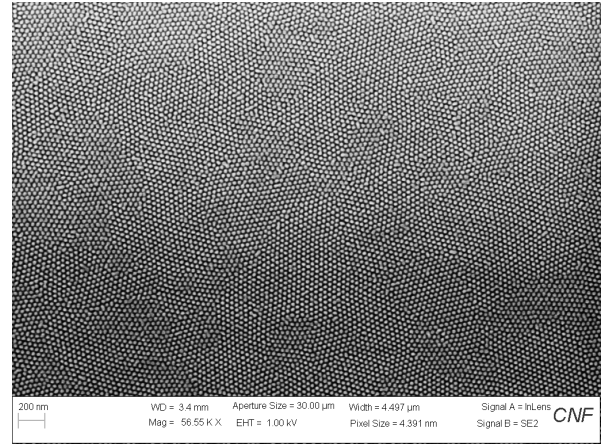}
\caption{SEM data of self-assembled graphene resonators}
\end{figure}

\begin{figure}[h]
\includegraphics[scale=0.4]{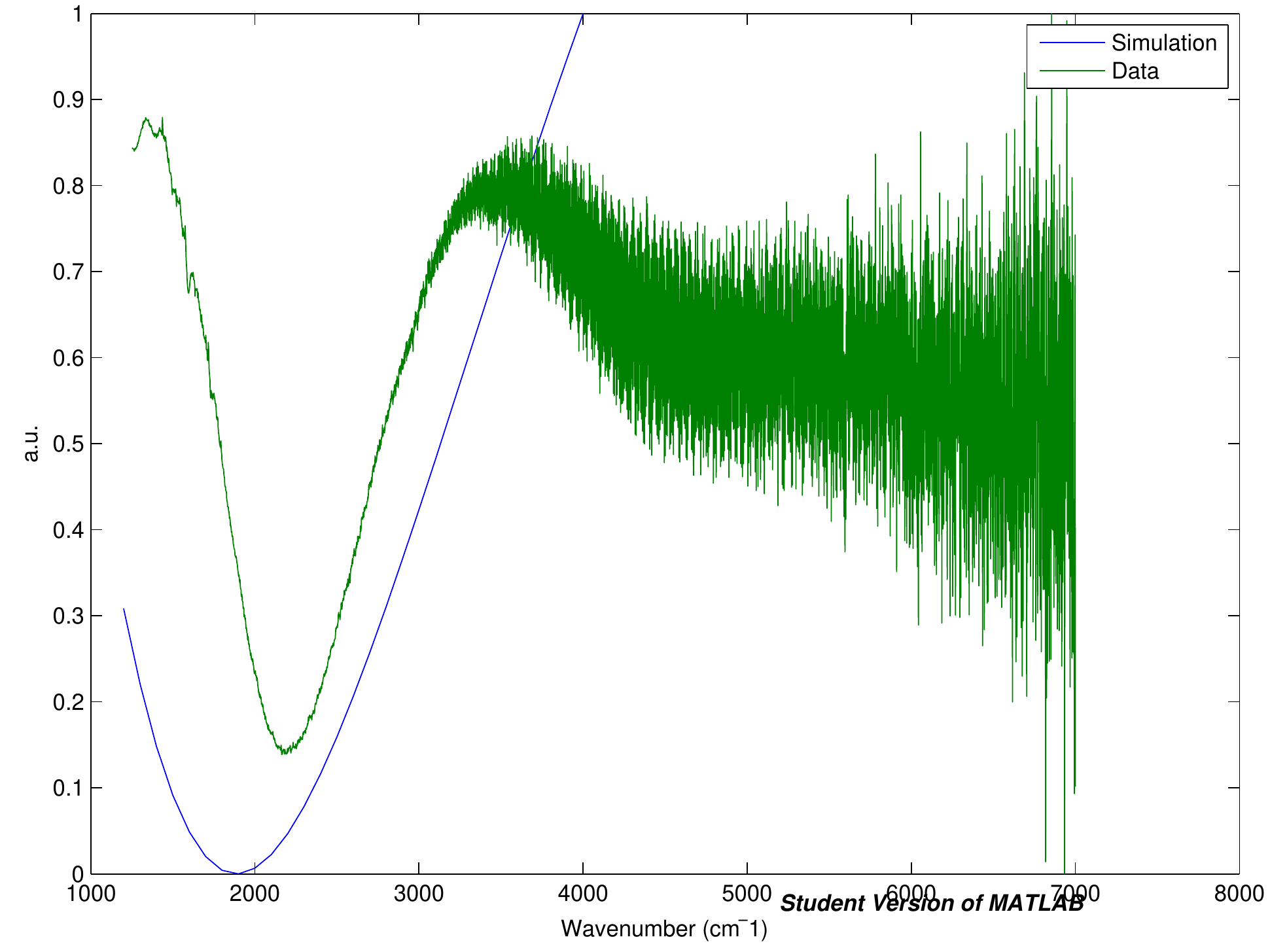}
\caption{Fit of model to FTIR transmission data. The y-axis shows the transmission through the sample and the x-axis is the wavenumber.}\label{fig:trans}
\end{figure}

We model a single plasmonic resonator following the approach in our previous paper\cite{Jared2013}. Starting from \begin{multline}\frac{\partial^{2}K_{i}(\vec{r})}{\partial t^{2}}+\frac{\partial K_{i}(\vec{r})}{\partial t}=\frac{\sigma(\omega=0)(\vec{r})}{\tau}[\frac{\partial E_{inc,i}(\vec{r})}{\partial t}\\-\frac{1}{\epsilon_{avg}}\int d\vec{r}'f_{i,j}(\vec{r}-\vec{r}')K_{j}(\vec{r}')] \end{multline} where K is the current density and $f_{ij}(\vec{r}-\vec{r}')=\partial_{i}\partial_{j}\frac{1}{|\vec{r}-\vec{r}'|}$. For the case of dots we have $K_{i}(\vec{r})=s(\vec{r})\frac{\sigma_{0}}{1-i\omega\tau}[E_{inc,i}(\vec{r})+\frac{1}{i\omega\epsilon_{avg}}\int d\vec{r}'f_{ij}(\vec{r}-\vec{r}')K_{j}(\vec{r}')]$ where $s(\vec{r})$ is a shape function which is 1 where graphene is present and 0 otherwise. Neglecting magnetic fields, we can represent $\vec{K}$ using a scalar potential $\nabla\phi$ where $\phi$ is related to the electrostatic potential via the constitutive relation. Rearranging we obtain \begin{multline}\int d\vec{r}'[\frac{1-i\omega\tau}{\sigma_{0}}\delta_{ij}\delta(\vec{r}-\vec{r}')-\frac{1}{i\omega\epsilon_{avg}}s(\vec{r})f_{ij}(\vec{r}-\vec{r}')]K_{j}(\vec{r'})\\=s(\vec{r})E_{inc,i}(\vec{r})\end{multline}. We can expand $\vec{K}$ in terms of a set of basis functions $\vec{b}_{\alpha}(\vec{r})$ where we let each basis function belong to a single resonator such that the bases on different dots are orthogonal. Then Bubnov-Galerkin discretization of the above equation gives $[\frac{1-i\omega\tau}{\sigma_{0}}\delta_{\alpha}{\beta}-\frac{1}{i\omega\epsilon_{avg}}F_{\alpha\beta}]K_{\alpha}=E_{inc}^{\alpha}$ where \begin{equation}F_{\alpha\beta}=\int d\vec{r}\int d\vec{r}'s(\vec{r})b_{\alpha i}^{*}(\vec{r})f_{ij}(\vec{r}-\vec{r}')b_{\beta j}(\vec{r}')s(\vec{r}')\end{equation} Since our structure is a disk of radius R, we choose $\vec{b}_{mn}(\vec{r})=c_{mn}\nabla(J_{m}(\frac{y_{mn}r}{R})e^{im\phi})$ and applying orthonormality to $\int d\vec{r}s(\vec{r})\vec{b}_{nm}^{*}\vec{b}_{m'n'}(\vec{r})$ we have $c_{mn}=\frac{1}{\sqrt{\pi (y_{mn}^{2}-m^{2})J_{m}^{2}(y_{mn})}}$. For a unit incident polarized plane wave $E_{inc}^{mn}=\int d\vec{r}s(\vec{r})\hat{x}\cdot\vec{b}_{mn}(\vec{r}) = \pi Rc_{mn}\delta_{|m|1}J_{m}(y_{mn})$. Returning to $F_{\alpha\beta}$, we take the fourier transform of $\frac{1}{|\vec{r}-\vec{r}'|}$ to get
\begin{multline}
F_{\alpha\beta}=\int \frac{d\vec{G}}{2\pi G}[\int d\vec{r}e^{i\vec{G}\vec{r}}s(\vec{r})\delta_{i}b_{\alpha i}(\vec{r})]^{*} \\ [\int d\vec{r}'e^{i\vec{G}\vec{r}'}s(\vec{r})\delta_{j}b_{\beta j}(\vec{r}')]
\end{multline} 
Then substituting back into equation 1 and plotting the eigenvalues vs wavenumber we obtain the plot in Fig.~\ref{fig:trans}.

Since broadening is important for many applications for example biosensing\cite{biosensor} and emission studies\cite{emmission} of the plasmons, we would like to characterize the broadening observed. We distinguish between process specific and process independent broadening. Process independent broadening include processes such as electron phonon coupling\cite{phononscattering}, electron-electron scattering\cite{DasSarma} and defects in the material itself\cite{defect}. Whereas process dependent broadening here refers to broadening due to edge roughness\cite{Debdeep}, doping variations, coupling between neighbouring resonators\cite{Pari2014},size distribution of the individual resonators, defects and grain boundaries\cite{BCPGrain} in the self assembled block copolymer film.

We first examine the disorder due to defects and grain boundaries in our self assembled block copolymer film. The presence of impurities on the surface of the SiO$_{2}$ layer may generate nonuniformities in the surface energy as well we block the adhesion of the random copolymer brush which gives rise to defects in the block copolymer film. In order to mitigate this effect we spun Acetone and Isopropyl Alcohol on the SiO$_{2}$ film before applying to random copolymer brush. Due to the size of the polymer chains there is limited mobility even at high annealing temperatures and this causes the formation of grains and grain boundaries. In order to account for the disorder in our film we extract the positions of the dots from our SEM data and set the $s(\vec{r})$ function in our model to 1 where graphene is present and 0 otherwise assuming dots of uniform radii. We perform simulation of our resonators assuming uniform doping and a block copolymer film with perfect translation and orientation correlation and observe that for such an array of resonators, the broadening is 10 cm$^{-1}$ while for a sample with our disorder again assuming uniform doping we have a broadening of 57 cm$^{-1}$. 

Next we examine the effect of coupling between neighbouring dots. This is a concern in our arrays because the evanescent tail from each resonator is able to couple to a neighbouring resonator due to their proximity and cause a shift in the resonance frequency as well as a coupled mode shape. Here we consider the interaction between cylindrical functions $f_{\vec{r_{1}}m_{1}(\vec{r})}$ and $f_{\vec{r_{2}}m_{2}(\vec{r})}$. Define $\tilde{f}_{m}(G) = \int rdr f_{m}(r)J_{m}(Gr)$ to be the Bessel transform of the radial part of the cylindrical function. The 2D Fourier transform of the cylindrical function centered at $\vec{r}_{1}$ is the same as before except for an overall phase factor $e^{i\vec{G}\cdot\vec{r}_{1}}$, therefore \begin{multline}F_{\vec{r}_{1}m_{1},\vec{r}_{2}m_{2}} \\ = (2\pi)^{2}e^{i(m_{2}-m_{1})\phi_{21}}\int_{0}^{\infty}\frac{d\vec{G}}{2\pi G}\tilde{f}_{\vec{r}_{1}m1}^{*}(\vec{G})\tilde{f}_{\vec{r}_{2}m2}^{*}(\vec{G})\\=(2\pi)^{2}e^{i(m_{2}-m_{1})\phi_{21}}\int_{0}^{\infty}dG\tilde{f}_{m_{1}}^{*}(G)\tilde{f}_{m_{2}}(G)J_{m_{1}-m_{2}}(Gr_{21})\end{multline}. Assuming uniform doping of the dots we see a shift in our resonance frequency of $129cm^{-1}$ wavenumber and a broadening of 10.3 cm$^{-1}$ vs 1.8 cm$^{-1}$ for the isolated dot. 

The third contribution to broadening is dopant distribution. Ideally a disk geometry should provide the minimal number of modes and hence the lowest broadening, the trade-off is that since the disk is an isolated structure we are unable to use electrostatic doping as any metals above or below will cause substantial damping, hence the use of chemical doping in our case, although surface doping may be a promising direction for future work. To account for the contribution of doping, we compare an array of resonators without defects or grain boundaries with a poisson doping distribution and with a uniform doping distribution. To account for this in our model we TODO. In the former case, the calculated broadening is 10 cm$^{-1}$ and in the latter case the broadening is 16 cm$^{-1}$. 

Lastly we would like to check for size dispersion of our individual resonators. Size dispersion in our devices can be caused by a nonuniform BCP film thickness but by positioning the 1 inch by 1 inch graphene sheet in the middle of the 50 cm by 50 cm quarter wafer and flooding the chip before spin coating we can minimize this effect. As mentioned previously we also maintain the cleanliness of the wafer backside so that the sample stays flat during plasma processing. We extract the size dispersion from our SEM data and assume a perfect dopant distribution as well as an array without defects or grain boundaries. Here we again employ the $s(\vec{r})$ function but we incorporate the actual size dispersion of the dots not just the positions. The broadening in our array assuming uniform doping is 60 cm$^{-1}$.

In order to conclude that the broadening is not due to our process we have to look at the cumulative effect of these individual contributions to broadening. Towards this end we have performed calculations incorporating combinations of these and we summarize them in table 1.
\begin{table}[ht]
\caption{Contributions to broadening from different sources}
\centering
\begin{tabular}{c c c c c c}
\hline\hline
Case & Disorder & Coupling & Dopant & Size & Combined \\ [0.5ex]
\hline
Broadening (cm$^{-1}$) & 10 & 10.3 & 10 & 60 & 55.8 \\ [1ex]
\hline
\end{tabular}
\label{table:nonlin}
\end{table}
\section{Conclusion}
In conclusion we have demonstrated self-assembled graphene plasmon resonators operating in the mid-IR. The area of our array is limited by the area of graphene available and the technique can produce even smaller resonators with higher frequency by increasing the PS to PMMA ratio. The broadening observed is independent of our process which suggest further studies of the intrinsic scattering mechanisms in graphene. Finally this method can be used to produce nanoscale dots in other 2D materials to study their electronic or plasmonic properties.
\bibliography{apstemplate}

\end{document}